\documentclass[twocolumn]{aastex631}
\usepackage{amsmath}
\usepackage{xspace}

\usepackage{multirow}

\newcommand{\cxo}{{\it Chandra}}

\newcommand{\revI}[1]{{#1}}
\newcommand{\revII}[1]{{#1}}
\newcommand{\revIII}[1]{{#1}}

\begin{document}

\title{{\em Chandra} Rules Out Super-Eddington Accretion \revII{Models} For Little Red Dots}

\correspondingauthor{Andrea Sacchi}\email{andrea.sacchi@cfa.harvard.edu}
\author[0000-0002-7295-5661]{Andrea Sacchi}
\author[0000-0003-0573-7733]{\'Akos Bogd\'an}
\affiliation{Center for Astrophysics $\vert$ Harvard \& Smithsonian, 60 Garden Street, Cambridge, MA 20138, USA}







\begin{abstract}
One of the most puzzling discoveries by \textit{JWST} is the population of high-redshift, red, and compact galaxies dubbed little red dots (LRDs). Based on broad-line diagnostics, these galaxies have been argued to host accreting  $10^7-10^8$~M$_\odot$ supermassive black holes (SMBHs), a claim with crucial consequences for our understanding of how the first black holes form and grow over cosmic time. A key feature of LRDs is their extreme X-ray weakness: analyses of individual and stacked sources have yielded non-detections or only tentative, inconclusive X-ray signals, except for a handful of individual cases. \revI{Although high obscuration is the most straightforward way to explain the X-ray weakness of LRDs, \textit{JWST} rest-frame optical/UV spectra initially argued against the presence of Compton-thick gas clouds. Instead, several authors have proposed that LRDs are intrinsically X-ray weak due to super-Eddington accretion rates.}
In this work, we \revI{observationally test these tailored models} by stacking X-ray data for 55 LRDs in the {\em Chandra} Deep Field South, accumulating a total exposure time of nearly $400$~Ms. Despite reaching unprecedented X-ray depths, our stack still yields a non-detection. The corresponding upper limits are deep enough to rule out \revII{current super-Eddington accretion models}, and are compatible only with extremely high levels of obscuration ($N_\textup{H}\gtrsim10^{25}$~cm$^{-2}$). To explain the X-ray weakness of LRDs, we therefore speculate that the SMBHs in these systems are neither as massive nor as luminous as currently believed. 
\end{abstract}



\section{Introduction} \label{sec:intro}

LRDs are high-redshift compact, red galaxies independently discovered by {\em JWST} \citep{harikane23,kokorev24,greene24,matthee24,kocevski24,labbe25}. These galaxies emerge at $z\approx8$, peak around $z\approx6$, and rapidly disappears at $z\approx4$ \citep{kocevski24}. The origin of their emission is still debated, with the two main interpretations being either intense star formation \citep[e.g.][]{baggen24} or accretion onto $10^7-10^8$~M$_\odot$ supermassive black holes (SMBHs) \citep[e.g.][]{kocevski23,greene24}. 

If LRDs host very massive SMBHs \citep{kocevski23,greene24}, this has important implications for our understanding of SMBH origins and growth. 
Such masses imply either that these SMBHs arose from ``heavy'' seeds  \citep{lodato06}, or from ``light'' seeds that grew at super-Eddington rates \citep{bromm03}. These SMBHs are also over-massive with respect to their host galaxies. While the masses of SMBHs are about $0.2\%$ of their host galaxies in the local universe \citep{magorrian98,reines15}, the mass of SMBHs in LRDs can be as high as  $10\%-100\%$ of their host galaxies \citep{harikane23,bogdan24,greene24,kocevski24,kovacs24}. 

If LRDs are unobscured AGN with an abundance of $10^{-5}\sim10^{-4}$~Mpc$^{-3}$~mag$^{-1}$ at redshift $z\approx5$ \citep{greene24}, their combined X-ray output would be substantial. In fact, if their spectral energy distribution (SED) is the same as local type 1 (unobscured) AGN, their integrated X-ray emission would exceed the observed X-ray background by an order of magnitude \citep{padmanabhan23}. This issue can be resolved by the X-ray weakness of LRDs. Safe for a handful of sources \citep{kocevski23,kocevski24}, LRDs are not detected in the X-ray band \citep{kocevski24,matthee24}, not even when stacking approaches are attempted \citep{ananna24,maiolino25}. The only tentative X-ray detection of LRDs, obtained by stacking objects for a total \cxo\ exposure of $\approx40$~Ms \citep{yue24}, implies optical-to-X-ray ratios that are orders of magnitude above those measured in ``typical'' AGN \citep{lusso10,sacchi22}.

\revI{Heavy obscuration is the most natural way to explain the X-ray weakness of LRDs. However, it requires some extreme parameters, such as column densities in excess of $\log N_\textup{H}\gtrsim24$ and near unity covering factors \citep{maiolino25}. Such a geometry is disfavored by {\em JWST} spectroscopic evidence, which reveals broad lines in the rest-frame optical/UV, thereby suggesting that we have a relatively unobscured view of the broad line regions of LRDs \citep[e.g.,][]{greene24}.} To alleviate this, several authors have proposed that LRDs accrete at super-Eddington rates, resulting in intrinsically X-ray-weak SEDs \citep{madau24,inayoshi24,pacucci24b,lambrides24}. In this picture, the lack of X-rays simply reflects the very steep spectral shape compared to more canonical (a few percent Eddington) accretion states.

To test the super-Eddington scenario, we analyze a sample of 55 LRDs with deep available \cxo\ data. \cxo\ is the best-suited instrument to search for faint X-ray emission of high-redshift sources, owing to its unmatched angular resolution and low and stable background. Given that all of the sources we considered lay in the deepest \cxo\ field, i.e.\ \cxo\ Deep Field South (CDF-S), the total exposure time, stacking the data from all of the considered sources, amounts to $\approx400$~Ms, allowing us to reach unprecedented sensitivity ($\approx4\times10^{-18}$~erg/s/cm$^2$). 

Although previous works were able to assess the X-ray weakness of LRDs, they could not reach the sensitivity required to probe the mechanisms at play. \citet{ananna24} analyzed \cxo\ data of LRDs lensed by the Abell 2744 cluster. They found that no source is individually detected, and neither is the stacked sample. The authors conclude that, based on the X-ray upper limits they obtained, the masses of the SMBHs hosted in LRDs cannot exceed a few $10^6$~M$_\odot$, assuming a standard unobscured AGN SED accreting at some percent of their Eddington rate. \citet{yue24} stacked 34 spectroscopically confirmed LRDs, confirmed their X-ray weakness, and obtained a tentative detection, favoring the AGN interpretation for the emission of LRDs.  

The present work surpasses both of these analyses: with respect to \citet{ananna24}, we can achieve better sensitivity as we do not have to take into account the bright foreground emission of a galaxy cluster; with respect to \citet{yue24}, we have more objects with much deeper X-ray images, allowing us to account ten times longer total exposure time. \revI{Both these works assessed the X-ray weakness of LRDs, and triggered the scientific community to propose different models to explain this feature.  Here, exploiting our significantly more sensitive dataset, we directly probe these theoretical models.}

This paper is organized as follows. In Section \ref{sec:xray} we describe our procedure to analyze (stack) the X-ray data; in Section \ref{sec:res} we present our results; in Section \ref{sec:disc} we discuss the implications of our X-ray non-detections and discuss the compatibility of our results with current models of X-ray weakness, and we finally draw our conclusions in Section \ref{sec:conc}. In this work we adopt a flat $\Lambda$CDM cosmology with $H_0=70$~km/s/Mpc and $\Omega_M=0.3$.

\section{Data analysis}\label{sec:xray}

\begin{figure*}[ht!]
\centering
\includegraphics[width = 0.90\textwidth]{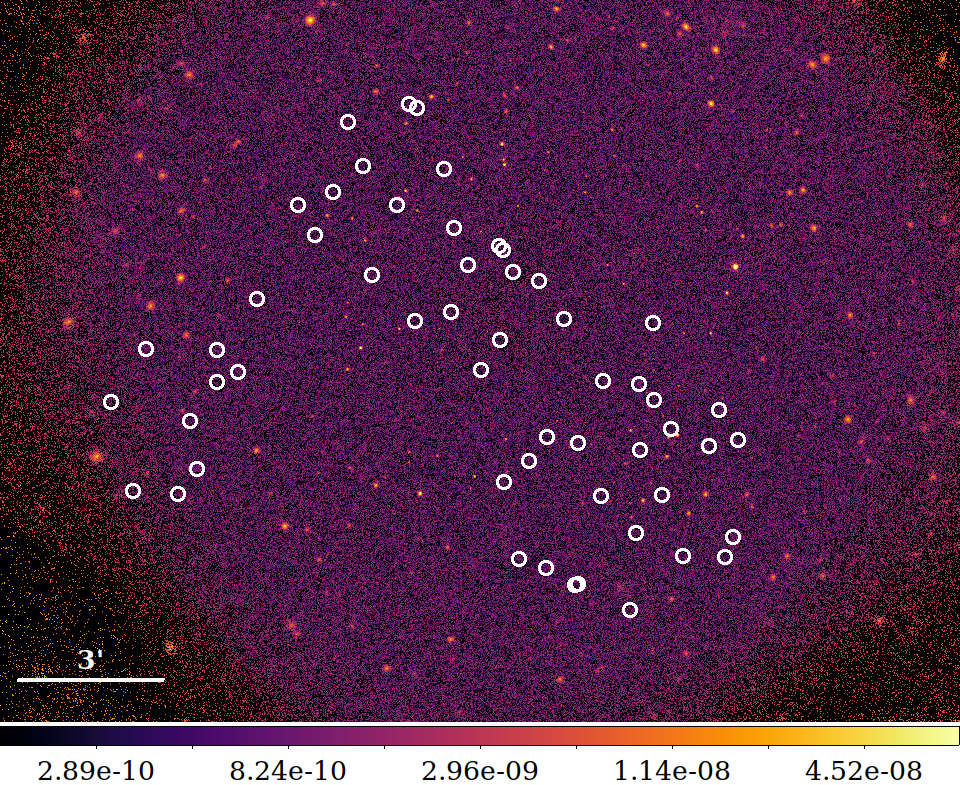}
\caption{Exposure corrected \cxo\ image in the $0.3-7$~keV X-ray band of the CDF-S, for which $\approx7$~Ms of total exposure time is available, making it the deepest \cxo\ dataset available to date. The location of each of the 55 LRDs considered in this work is indicated by a white circle with fixed $8\arcsec$ radius for illustration purposes. \label{fig:xray_img}}
\end{figure*}

\subsection{The sample of LRDs}

Our parent sample is composed of LRDs identified in the {\em JWST} Advanced Deep Extragalactic Survey (JADES) \citep{eisenstein23} and the Next Generation Deep Extragalactic Exploratory Public (NGDEEP) \citep{bagley24} surveys by \citet{kocevski24}. These 56 sources span from redshift $z\approx3$ to $z\approx8$, \revII{however, the bulk of the population has redshifts in the range $z=4.5-7.5$,} with a mean redshift of $z\approx6$. \revI{For most of the LRDs in our sample only photometric redshifts are available. However, the reliability of \textit{JWST}-derived photometric redshift has been demonstrated by multiple spectroscopic follow-ups  \citep[e.g.][]{goulding23, greene24, napolitano24}. In addition, the conclusions presented in this work are based on the average LRD population and are hence insensitive to small redshift uncertainties associated with individual galaxies. }

The individual X-ray emission of these sources is discussed in \citet{kocevski24}. Only one source, JADES 21925, at redshift $z=3.1$ is X-ray detected, and hence we exclude this source from further analysis. Hereafter, we only address the 55 LRDs that are not individually X-ray detected.

We selected these sources because they lie entirely within the footprint of CDF-S \citep{giacconi02,luo17}, which provides the deepest X-ray observation ever taken with \cxo, totaling $\approx$7~Ms exposure. Furthermore, most of these observations were taken before 2010, prior to the substantial loss of soft-band sensitivity from the molecular contamination \citep{plucinsky18,plucinsky22}. 

\subsection{X-ray stacking procedure}

We downloaded 104 \cxo\ observation of the CDF-S from the \cxo\ data archive\footnote{\url{https://cda.harvard.edu/chaser/mainEntry.do}} and performed the data analysis with the Chandra Interactive Analysis of Observations software package \citep[CIAO, v.4.17][]{fruscione06} and the CALDB 4.12.0 release of the calibration files. The first step of the analysis was reprocessing the data using \texttt{chandra\_repro} tool. To improve the absolute astrometry of the merged data, we used the \texttt{fine\_astro} routine, which cross-matches \cxo\ point sources with other higher precision catalogs, in our case the U.~S.~Naval Observatory catalog \citep[USNO-A2.0][]{monet98}. The astrometric-corrected and reprojected event files were then merged with the \texttt{merge\_obs} routine, totaling $\approx7$~Ms of exposure time. The same tool also produced point-spread function (PSF) and exposure maps, created assuming a powerlaw spectral profile with photon index $\Gamma=2$ and line-of-sight Galactic column density $N_\textup{H}=6.7\times10^{19}$~cm$^{-2}$.

To avoid contamination from other foreground sources, all point-like sources were identified with the \texttt{wavdetect} tool (with wavelet scales equal to 1.414, 2, 2.828, 4, 5.636, and 8) and accordingly removed. The exposure time for each single source amounts to $\approx7$~Ms, implying that the exposure for the stacked sample amounts to 390~Ms. Figure \ref{fig:xray_img} shows the merged \cxo\ view of CDF-S in the full ($0.3-7$~keV) band with the location of each LRD indicated.
Source counts were extracted from circular regions centered on the LRD positions as identified by {\em JWST}, and background counts from annular regions with the \texttt{dmextract} tool. The radius of the extraction region was chosen to correspond to the $R_{80}$ (the radius that includes 80\% of the source counts) at the location of each source, and the background regions had radii of ($1.5-3)R_{80}$. We note that the choice of adopting $R_{80}$ (instead of the more commonly used $R_{90}$) does not affect the significance of the (non-)detection, but contributes to lowering the background, in particular for the more off-axis sources. For our 55 sources, \revI{$R_{80}$ is in the range $0\farcs6-6"$, with the median source extraction region being $R_{80} = 2\arcsec$}. 

\section{Results}\label{sec:res}

\revI{By co-adding the X-ray photons for all 55 LRDs, we measure 1,444 counts in the soft ($0.3-2$~keV) and 3,730 counts in the hard ($2-7$~keV) band. Scaling the background to the same area results in 1,409 and 3,697 background counts in the soft and hard bands, respectively. Using Poisson statistics, this yields $35\pm38$ net counts in the soft band and $33\pm61$ in the hard band, corresponding to clear non-detections.} In the absence of detections, we derive upper limits on the flux of the stacked LRD population. \revI{To convert the count rates to fluxes, we utilize the \textit{Chandra} exposure maps, which correct for the detector response, the molecular contamination, and vignetting effects.  We thus obtain} $3\sigma$ upper limits of $<4.7\times10^{-18}$~erg/s/cm$^2$ in the soft and $<1.3\times10^{-17}$~erg/s/cm$^2$ in the hard band. We note that the particular choice of photon index adopted to convert the count rates into fluxes affects these latter by only $\approx20\%$, leaving our main results unchanged.

We convert our flux limits into bolometric luminosities by adopting standard bolometric corrections \citep[$L_\textup{bol}/L_\textup{X}=16.7$,][]{lusso12,duras20} \revI{and the mean redshift of the sources in our sample ($z=6$)}. We obtain an upper limit of $L_\textup{bol}< 3\times10^{43}$~erg/s, which is more than an order of magnitude lower than the average bolometric luminosities of LRDs inferred from {\em JWST} observations \citep[$L_\textup{bol}\approx5\times10^{44}$~erg/s,][]{harikane23,kocevski24}. 

Our results are in line with those obtained by previous works: stacking the X-ray data of individually non-detected LRDs results in a non-detection \citep{ananna24,maiolino25}. However, we do not confirm the tentative detection of \citet{yue24}.

\begin{figure*}[ht!]
\centering
\includegraphics[width = 0.90\textwidth]{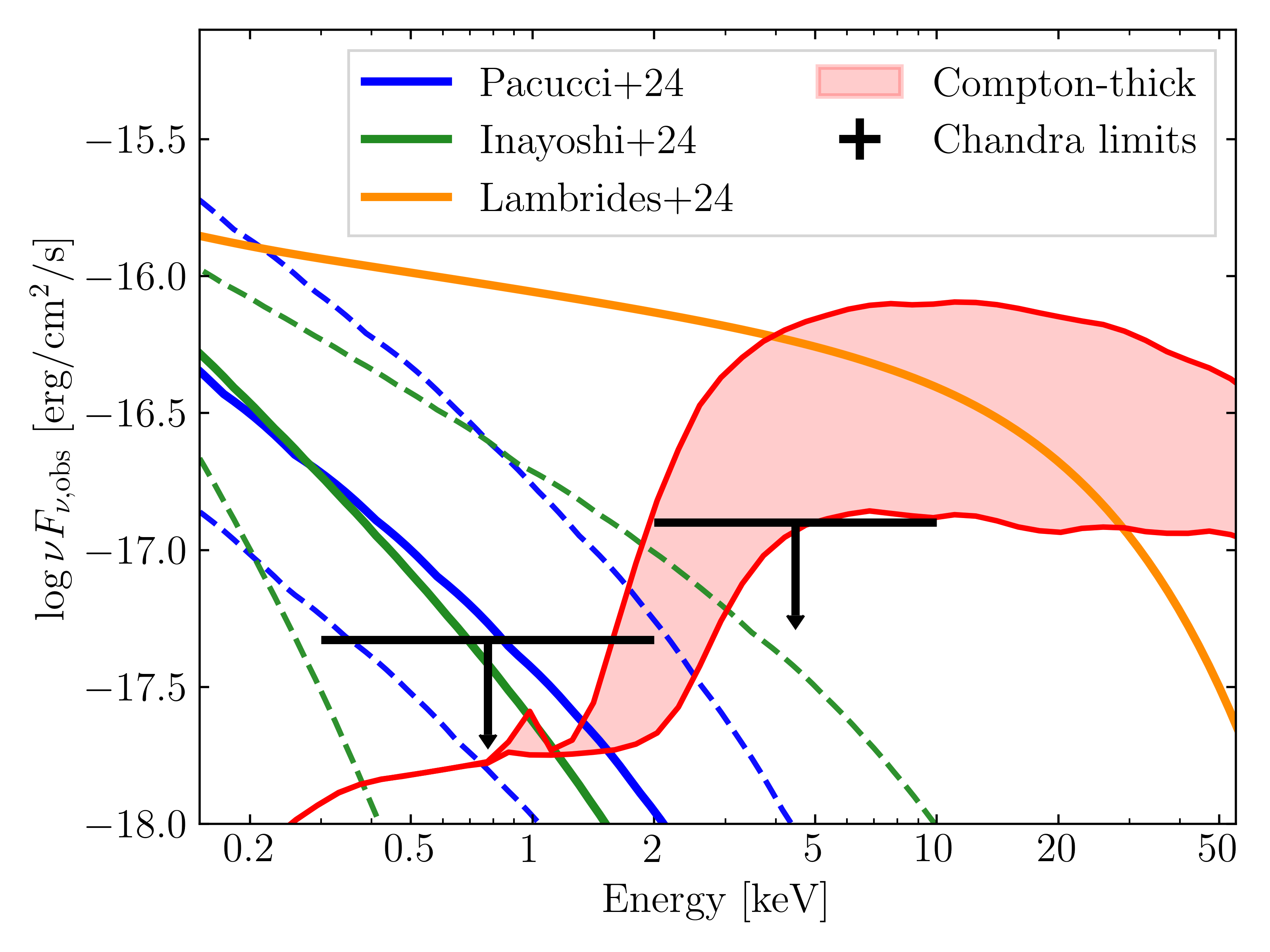}
\caption{Comparison between the upper limits on the X-ray emission of LRDs derived in this work and mock spectra corresponding to different scenarios of X-ray weakness. Intrinsic X-ray weakness from super-Eddington accretion from \citet{pacucci24b} and \citet{inayoshi24} are reported in blue and green, respectively. For \citet{pacucci24b}, the solid line indicates the mean expected flux, the dashed lines encompass a region spanning the viewing angles range $\theta=30^\circ-80^\circ$. For \citet{inayoshi24}, the solid line indicates an Eddington ratio $\lambda_\textup{Edd}=1$, and the dashed lines encompass a region which spans the range of Eddington ratios $\lambda_\textup{Edd}=0.3-3$. In orange is shown the model for super-Eddington accretion \texttt{agnslim}, assumed by \citet{lambrides24}. In red is shown a standard Compton-thick spectrum \citep[\texttt{borus02},][]{balokovic19}, the solid lines corresponding to column densities of $N_\textup{H}=10^{24.5-25}$~cm$^{-2}$. \label{fig:spec}}
\end{figure*}

\section{Discussion}\label{sec:disc}

Here we discuss possible explanations for the extreme X-ray weakness we inferred for the LRDs in our sample. We discuss three potential scenarios:
\begin{itemize}
    \item[-] The sample may include non-AGN contaminants, hence the signal from the accreting SMBHs is diluted by the presence of spurious sources;
    \item[-] The SED of the LRDs is that of a standard type 1 AGN, but the X-ray emission is heavily obscured;
    \item[-] The SED of the LRDs is not that of standard type 1 AGN, and their X-ray emission is intrinsically weak. 
\end{itemize}
    
\subsection{Non-AGN contaminants}
Some of the objects in our sample could be high-redshift galaxies without active SMBH accretion, or foreground interlopers such as brown dwarfs. However, to explain the lack of X-ray emission of LRDs invoking uniquely non-AGN contaminants, one needs to assume that the majority of the considered objects are spurious sources \citep{maiolino25}. This scenario has already been ruled out by {\em JWST} spectroscopic follow-ups of early compilations of LRD samples, which confirmed the presence of broad $H_\alpha$ emission lines for most of them, reinforcing the AGN interpretation \citep{kocevski23}. \citet{greene24} followed up photometrically-selected LRDs and found that $\approx60\%$ show broad $H_\alpha$ emission lines, for a small fraction ($\approx20\%$) the line diagnostic was inconclusive, and a final $\approx20\%$ was non-AGN contaminants, i.e.\ brown dwarfs. Our sample of LRDs, however, was carefully selected to avoid the presence of such objects based on their infrared colors \citep{kocevski24}, hence we do not expect this last fraction to be as significant. Nonetheless, in the rest of the discussion, we contemplate the possibility that $20\%$ of the sources may be non-AGN contaminants.

\subsection{High obscuration}

High column densities could be responsible for the X-ray weakness of LRDs. As already argued in previous studies \citep{ananna24,maiolino25}, to hide the X-ray emission at the measured levels, column densities in excess of $>10^{24}$~cm$^{-2}$ are needed, implying a Compton-thick regime. Residual X-ray emission is expected from reflection on the material surrounding and obscuring the SMBHs. This scenario needs further tuning to account for the multi-wavelength features of LRDs (e.g.\ the absence of dust and the presence of broad emission lines \citealt{maiolino25}), but a discussion of these aspects is beyond the scope of this work. 

To test if Compton-thickness is compatible with the deep \cxo\ upper limits derived from our stacking analysis (Section~\ref{sec:res}), we compare these with mock spectra of Compton-thick AGN with different values of column densities. The mock spectra were created with the \texttt{borus02} model \citep{balokovic19}. The normalization of the transmitted component was set assuming a bolometric luminosity of $L_\textup{bol}=5\times10^{44}$~erg/s, and the standard bolometric correction of $k_\textup{bol}=16.7$ \citep{duras20}. We assumed a covering factor $\mathrm{CF}=1$ and an inclination angle of $\cos\theta=0.05$. The model is also shifted to the mean redshift of our sample. Figure~\ref{fig:spec} shows the comparison between the mock spectra and the upper limits we derived. The solid red lines correspond to column densities of $N_\textup{H}=10^{24.5-25}$~cm$^{-2}$. When compared to the \cxo\ upper limits, reported in black, one can see that only extremely high values of column densities $N_\textup{H}\gtrsim10^{25}$~cm$^{-2}$ are permitted. \revI{Such extreme values of $N_\textup{H}$ are in contrast with results obtained from high-resolution  {\em JWST} spectroscopy (Section~\ref{sec:intro}). However, \citet{maiolino25} show that if the obscuration originates in the same clouds responsible for the broad line emission, and if these clouds are located within the dust-sublimation region, and are hence dust-free, then this scenario can provide Compton-thick columns without altering the optical/UV appearance of LRDs.}

\subsection{Intrinsic X-ray weakness}

Intrinsic X-ray weakness, motivated by accretion rates at or above the Eddington limit, has been invoked by multiple authors to explain the X-ray non-detections of LRDs \citep{inayoshi24,madau24,lambrides24,pacucci24b,naidu25}. \revI{In the local universe, AGN accreting at high Eddington fractions usually exhibit softer and less luminous X-ray spectra \revI{\citep{vasudevan07,tortosa22,tortosa23}}. However, these sources are bright enough that even one such source in our sample would be individually detected. For example, I~Zwicky~1, with $L_{\rm 0.3-10keV}\approx10^{44}$~erg/s \citep{gallo07}, if placed at $z=6$, would produce 200 net counts (a $\sim12\sigma$ detection) in the CDF-S field. Consequently, tailored super-Eddington accreting models have been developed specifically to reproduce the extreme X-ray weakness observed in LRDs.}

As prototypes for super-Eddington accretion models, we adopt the spectral profiles presented by \citet{pacucci24b} and \citet{inayoshi24}. Using general relativistic radiation magnetohydrodynamics (GRRMHD) simulations, \citet{pacucci24b} obtained a family of X-ray spectral profiles of SMBHs accreting at super-Eddington rates, observed at different viewing angles. \citet{inayoshi24} performed a similar exercise via an analytical approach, spanning different mass accretion rates. 

Both families of spectra have bolometric luminosities compatible with those inferred for LRDs, and we redshifted them to the mean redshift of the sources in our sample ($z=6$). \revII{We adopted the models presented in Fig.\ 4 (bottom-left panel) of \citet{pacucci24b}, corresponding the X-ray weak case}, and in Fig.\ 3 (left panel) of \citet{inayoshi24}, corresponding to the case of an AGN powered by accretion onto a $10^7$~M$_\odot$ SMBH. 

In Fig.~\ref{fig:spec}, we compare these two models with our upper limits. In blue is the model by \citet{pacucci24b}: the dashed lines encompass a region which spans the range of viewing angles $\theta=30^\circ-80^\circ$, and the solid line indicates the average spectrum. In green is the family of SEDs derived by \citet{inayoshi24}: the dashed lines encompass a region corresponding to a range of Eddington rates $\lambda_\textup{Edd}=0.3-3$ and the solid line indicates $\lambda_\textup{Edd}=1$ ($\lambda_\textup{Edd}$ is mass accretion rate in units of Eddington rate). Both models, although derived in different and independent ways, show similar profiles. This is a common trait for super-Eddington accretion models; they all predict extremely soft X-ray emission. Indeed, when parametrized with a power law profile, typical photon indices are $\Gamma\gtrsim3$. 

\revI{Figure~\ref{fig:spec} summarizes the results of our stacking analysis. \cxo\ upper limits rule out all considered models of super-Eddington accretion. For the \citet{pacucci24b} models, not only is the average spectrum ruled out, but even the most extreme viewing angles, as they all lie above the upper limits derived for our sample. \revII{Even in the implausible scenario in which all of the SMBHs hosted in LRDs are seen edge-on, the resulting X-ray emission would be detected by \cxo\ with a $>3\sigma$ significance}. For \citet{inayoshi24}, only the most extreme values of accretion rate are not ruled out ($\lambda_\textup{Edd}\gtrsim3$), as the spectrum is so steep that it falls outside the soft X-ray band.}

We note that even assuming that only $80\%$ of the LRDs in our sample actually host an accreting SMBH, it would not introduce a sufficiently large correction to reconcile the examined model with the \cxo\ upper limits. 

The super-Eddington SEDs presented here represent excellent prototypes for this family of models, and our conclusions also extend to those proposed by other authors, as all of them predict similar X-ray slopes and luminosities. For completeness, in Fig.~\ref{fig:spec} we also show (in orange) the super-Eddington accretion model adopted by \citet{lambrides24}. These authors did not derive an SED, but assumed one to provide seed photons to a photo-ionization model to reproduce the emission lines observed in LRDs spectra. \revII{The model assumed for super-Eddington accretion is a slim disk spectrum \citep[\texttt{agnslim},][]{kubota19}, that we set to have the same parameters as those described by \citet{lambrides24}.} The presence of a warm corona in this model, in addition to a slim accretion disk, generates a considerable amount of X-ray emission, exceeding by $1-1.5$ orders of magnitude the upper limits we derived, and it is hence ruled out.

\revI{Finally, beyond our stacking limits, the two X-ray detected LRDs exhibit spectra that are inconsistent with \revIII{the published} super-Eddington accretion \revIII{models, which predict photon indeces of $ \Gamma\gtrsim3$.} Indeed, both X-ray sources are well fit by power-law models with photon indices of $\Gamma\approx2$, which is typical in standard, unobscured AGN \citep{kocevski24}, further disfavoring the \revIII{proposed} super-Eddington accretion \revIII{models}.}

Based on this evidence, we conclude that present \cxo\ observations rule out \revII{\revIII{current} super-Eddington accretion \revIII{models} that aim to explain the population-wide X-ray faintness of LRDs. However, because stacking averages over the entire sample, we cannot exclude that a small subset of LRDs undergoes super-Eddington accretion.} \revIII{Similarly, we note that future theoretical work may produce revised super-Eddington models, whose predicted X-ray output would fall below our stacking limits.}

\subsection{Overestimated masses and bolometric luminosities}

Based on the evidence reported so far, \cxo\ data rule out \revIII{current models} of super-Eddington accretion, and are compatible only with extremely high levels of obscuration. 
To reconcile the X-ray upper limits with our current understanding of accretion onto SMBHs, it has been argued that the masses of the SMBHs hosted in LRDs are overestimated \citep{ananna24,maiolino25,rusakov25}. However, this alone cannot solve the puzzle of the multi-wavelength behavior of LRDs, as the proposed models are normalized on the bolometric luminosities rather than on the SMBHs masses. Hence, lowering the masses of the SMBHs will not affect the normalization of the models, but will simply increase the inferred accretion rates.

As full (multi-wavelength) SED-modeling for LRDs is not feasible, bolometric luminosities are inferred either from the luminosities of the broad $H_\alpha$ line or of the rest-frame optical/UV continuum, exploiting empirical relations derived from local unobscured AGN \citep[e.g.][]{stern12}. We argue that the underlying assumption that LRDs share the same SED of local unobscured AGN is fundamentally disproved by their X-ray weakness, and we conclude that the bolometric luminosities are overestimated.

If we assume lower bolometric luminosities, more modest column densities would be sufficient to obscure the X-ray emission from LRDs, with no need to invoke super-Eddington accretion. Although reducing the bolometric luminosities, and hence the SMBH masses, would make super-Eddington models compatible with our X-ray upper limits, such models would be unnecessary, as standard accretion regimes could fully explain the observed X-ray properties.

\section{Conclusions}\label{sec:conc}

In this work we explored the X-ray properties of 55 LRDs identified in the JADES and NGDEEP surveys. These objects are not individually X-ray detected, and also by stacking all of them, we obtained a non-detection, both in the soft and hard X-ray band. This result falls in line with those obtained by exploiting different datasets, but the $3\sigma$ upper limits we derived, based on a total exposure time of $\approx390$~Ms, are the deepest ever obtained. 

Leveraging this unprecedented sensitivity, we tested super-Eddington accretion models proposed to explain the X-ray weakness of LRDs. None of these models is compatible with our observations, ruling out \revIII{current} super-Eddington accretion \revIII{models}. Furthermore, since non-AGN contaminants are expected to be negligible, our X-ray results require extreme Compton thickness.

To relax this tension, we argue that both the bolometric luminosities and SMBH masses are overestimated, which has been suggested by several authors. If the bolometric luminosities are overestimated by an order of magnitude, much lower levels of obscuration can hide the X-ray emission from accreting SMBHs without invoking super-Eddington accretion.

\bigskip
\section*{Acknowledgments}

\'A.B. acknowledges support from the Smithsonian Institution and the Chandra Project through NASA contract NAS8-03060. \revI{The plot presented in Figure 2 was inspired by Mazzolari et al.\ (in preparation).} This research has made use of data obtained from the Chandra Data Archive and the Chandra Source Catalog, both provided by the Chandra X-ray Center (CXC). This paper employs a list of Chandra data sets, obtained by the Chandra X-ray Observatory, available at~\dataset[DOI: 10.25574/cdc.382]{https://doi.org/10.25574/cdc.382}.

\vspace{5mm}
\facility{CXO}
\software{Astropy \citep{astropy13,astropy18,astropy22}, Matplotlib \citep{matplotlib07}, NumPy \citep{numpy20}, \revI{DS9 \citep{ds9}}}

\bibliography{biblio}{}
\bibliographystyle{aasjournal}
\end{document}